\documentclass[aps,pra,twocolumn,floatfix,longbibliography,showpacs]{revtex4-2}

\usepackage{amsmath,amsfonts,amssymb,bm}
\usepackage{dcolumn}
\usepackage{braket}
\usepackage{tabularx}
\usepackage{graphicx}
\usepackage[update,prepend]{epstopdf}

\usepackage{dsfont}
\usepackage{pifont}

\newcommand{\iu}{\mathrm{i}\mkern1mu}



\usepackage{physics}
\usepackage{hyperref}
\usepackage[capitalize]{cleveref}
\usepackage{xcolor}

\usepackage{booktabs}

\begin{document}

\title{Elucidating the Control of Circular Dichroism in Ion Yield via Chirped Pulses with Purposeful Models}

\author{Leon A. Kerber}
\affiliation{Freie Universität Berlin, Dahlem Center for Complex Quantum Systems and Fachbereich Physik, D-14195 Berlin, Germany}
\author{Daniel M. Reich}
\affiliation{Freie Universität Berlin, Dahlem Center for Complex Quantum Systems and Fachbereich Physik, D-14195 Berlin, Germany}

\begin{abstract}
  We theoretically investigate circular dichroism in the ion yield following $1+1+1$ ionization of 3-methylcyclopentanone using femtosecond linearly chirped laser pulses, inspired by recent experiments by Das \emph{et al.}\ [Phys.~Chem.~Chem.~Phys.~\textbf{27}, 8043 (2025)].
  To this end, we numerically solve the time-dependent Schr\"odinger equation and evaluate the total population in the Rydberg states at the end of the second absorption step.
  The A-band transition in the first absorption step is treated using state-of-the-art quantum-chemical calculations, whereas the second absorption step is described via an effective model.
  Within our framework, we identify the interplay between the first and second absorption step as the key explanation for the experimentally observed chirp dependence of the anisotropy.
  By elucidating this mechanism for the chirp-enhanced signal, our findings contribute towards the development of improved control schemes for chiral molecules.
\end{abstract}

\maketitle

\section{Introduction}

The differential absorption of left- and right-circularly polarized light by optically active chiral molecules is known as circular dichroism (CD)~\cite{CDnakanishi}.
In contrast to absorption of unpolarized or linearly polarized light (ABS), which is dominated by electric-dipole transitions, CD arises from the interplay of electric and magnetic transition dipole moments.
As a result, the ratio between CD and ABS, the so-called anisotropy factor or $g$-factor~\cite{Kuhn1930,berova_application_2007}, is typically very small.
However, if the magnitudes of electric and magnetic dipole moments become comparable, the anisotropy factor can reach comparatively large values.
An important example is found in the first singlet electronic transition band (the A band) of ketones, where anisotropy factors of the order of $10\,\%$ are observed~\cite{CDnakanishi}.
At the same time, the theoretical description of this transition is challenging due to the coupling between electronic and vibrational degrees of freedom.
Specifically, the Franck–Condon approximation~\cite{franck_elementary_1926,condon_theory_1926,condon_nuclear_1928} breaks down and Herzberg–Teller contributions~\cite{herzberg_schwingungsstruktur_1933} must be taken into account~\cite{moffitt_optical_1959,weigang_vibrational_1965,atkins1997}.

A particularly well studied ketone is 3-methyl\-cyclo\-penta\-none (3MCP), which has been the subject of numerous theoretical and experimental investigations~\cite{boesl_von_grafenstein_circular_2006,li_linear_2006,bornschlegl_investigation_2007,loge_progress_2009,breunig_circular_2009,boesl_resonance-enhanced_2013,titze_laser_2014,ring_self-referencing_2021,horsch_circular_2011,das_control_2025,lin_vibronically_2008,feinleib_vapour-phase_1968,dekkers_optical_1976,kerber_anisotropy_2025,urry1968}.
Notably, it was the first molecule in which CD was measured via differential ion yields (CDIY)~\cite{boesl_von_grafenstein_circular_2006,li_linear_2006}.
CDIY has since been measured in several subsequent studies~\cite{boesl_resonance-enhanced_2013,loge_progress_2009,titze_laser_2014,horsch_circular_2011,breunig_circular_2009,bornschlegl_investigation_2007,ring_self-referencing_2021}.
In such experiments, ionization is typically achieved via resonance-enhanced multiphoton ionization (REMPI), where the molecule is excited through one or more resonant intermediate states before ionization occurs.

Further studies have investigated possibilities for control of the anisotropy in a $1+1+1$ ionization scheme~\cite{horsch_circular_2011,das_control_2025}, where the molecule is excited first to the A band and subsequently to one out of multiple Rydberg states before ionization, see \cref{fig:ionization_scheme}.
Specifically, it was found that the anisotropy factor is increased by linearly chirping femtosecond laser pulses at around $310\,\mathrm{nm}$.
While early studies reported that the enhancement of the anisotropy was essentially independent of the sign of the applied chirp~\cite{horsch_circular_2011}, in a recent experiment a clear asymmetry between positive and negative chirps was observed~\cite{das_control_2025}.
By establishing a theoretical understanding for this observation, we aim to move closer to improved control schemes that could further enhance the observed CDIY.

Theoretical studies so far have focused on the dependence of CD on laser pulse characteristics such as frequency~\cite{kroner_chiral_2011} and duration~\cite{horsch_circular_2011, kroner_chiral_2011}, or have described effects such as electric quadrupole interactions to permit coherent interference between excitation pathways~\cite{mondelo-martell_increasing_2022,mondelo-martell_correction_2023}.
These approaches, however, were typically limited by their neglection of the vibrational degree of freedom and second excitation step~\cite{kroner_chiral_2011,horsch_circular_2011,mondelo-martell_correction_2023,mondelo-martell_increasing_2022}.
We considerably extend the theoretical description by accounting for both --- the vibrational structure of the A band of 3MCP, based on state-of-the-art quantum-chemical calculations~\cite{kerber_anisotropy_2025} and the second absorption step.

The latter is considered crucial, because in resonance-enhanced two-photon absorption processes chirping affects the final state population by virtue of a sequential resonance mechanism~\cite{cao_intrapulse_1998,yakovlev_chirped_1998,balling_interference_1994}.
Moreover, such a mechanism can lead to preferential enhancement of some excitation pathways over others~\cite{krug_coherent_2009,plenge_chirped_2009}, depending on the sign and magnitude of the applied chirp.
Its relevance for 3MCP originates from the involved A-band transition due to the strong dependence of the anisotropy factor on the specific vibronic transition as a result of non-Condon effects~\cite{kerber_anisotropy_2025}.
However, quantum-chemical calculations of the relevant transition energies and matrix elements from the A state to the Rydberg states are challenging~\cite{casida_molecular_1998,goetz_theoretical_2017,smydke_using_2023,creutzberg_computing_2023}.
For this reason, we introduce two models to effectively describe the second absorption step.

We begin by developing a ``minimal model'' containing only a few non-interfering excitation pathways, which enables us to pinpoint the key mechanisms at the level of individual pathways.
We then mimic the manifold of involved Rydberg states and their vibrational substructures by randomly generating transition energies and matrix elements in what we call the ``randomized model''.
Remarkably, the resulting description proves to be robust with respect to the specific random realization, allowing us to draw fairly general conclusions.

The structure of this paper is as follows.
In \Cref{sec:theo_frame}, we present the theoretical framework.
In \Cref{sec:res}, we first analyze the results of the ``minimal model'' to uncover the fundamental mechanisms. Then, we examine the results of the ``randomized model'', comparing them with experimental observations.
Finally, \Cref{sec:conclusions} concludes.

\section{Theoretical Framework}\label{sec:theo_frame}

\begin{figure}[bt]
  \centerline{\includegraphics[width=8.6cm]{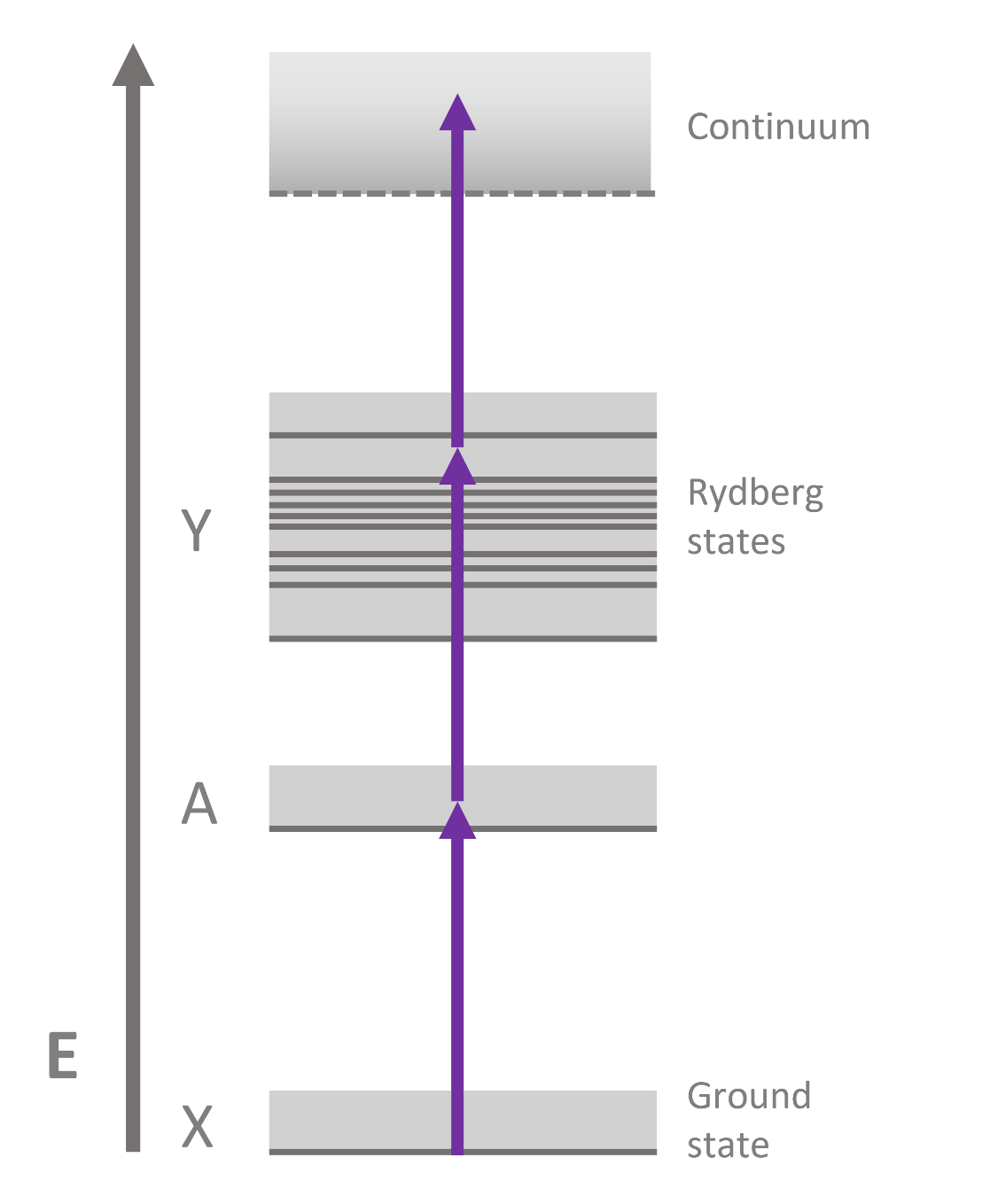}}
  \caption{The $1+1+1$ ionization scheme of 3MCP examined within this work.
    The absorption of a first photon excites the molecule to the A band.
    In a second absorption step, the molecule is promoted to one of ten Rydberg states~\cite{otoole_vacuum_1992}.
    Absorption of the third photon leads to ionization.
    The ionization step is treated effectively in our work.
  }\label{fig:ionization_scheme}
\end{figure}

CDIY is defined as the difference in ion yield obtained upon irradiation with left- and right-handed circularly polarized light.
The anisotropy factor $g$ is defined by normalizing the CDIY, $I_{\mathrm{CD}}=I_{\mathrm{left}}-I_{\mathrm{right}}$, with respect to the mean ion yield, $I_{\mathrm{ABS}}=\frac{1}{2}\qty(I_{\mathrm{left}}+I_{\mathrm{right}})$, that is,
\begin{equation}
  g=\frac{I_{\mathrm{CD}}}{I_{\mathrm{ABS}}}.
\end{equation}
Motivated by the experimental measurements of Ref.~\cite{das_control_2025}, we use laser pulses with central wavelengths between $308\,\mathrm{nm}$ and $311\,\mathrm{nm}$, which corresponds to the vicinity of the second peak in the absorption and anisotropy factor spectra of 3MCP~\cite{kerber_anisotropy_2025}.
The ionization scheme is illustrated in \Cref{fig:ionization_scheme}.

In the first absorption step, the molecule is promoted from its electronic ground state (X) to the first excited singlet state (A), which is approximately an electric-dipole-forbidden $n\rightarrow\pi^*$ transition within the carbonyl group.
Since this transition is only weakly allowed, the resulting $g$-factor depends on the specific vibrational states involved~\cite{kerber_anisotropy_2025}.
The second absorption step excites the molecule from the A state to one of several Rydberg-like electronic states (Y), whose vibrational manifolds overlap energetically.
In particular, ten Rydberg states (3s, 3p, 3d, and 4s) are plausible candidates for the final bound state~\cite{otoole_vacuum_1992}.
The third photon then ionizes the molecule.

\subsection{Hamiltonian and Observable}

The Hamiltonian of the system consists of the field-free molecular Hamiltonian $\hat{H}_0$ together with the electric- and magnetic-dipole interactions with the laser field.
In the semi-classical framework, assuming the electromagnetic fields $\vb*{E}\qty(t)$ and $\vb*{B}\qty(t)$ to propagate along the z-direction, this yields
\begin{align}
  \hat{H}\qty(t)&=\hat{H}_0-\vb*{E}\qty(t)\cdot \hat{\vb*{\mu}}-\vb*{B}\qty(t)\cdot \hat{\vb*{m}} \notag \\
  &=\hat{H}_0-\vb*{E}\qty(t)\cdot \hat{\vb*{\mu}}-\frac{1}{c}\qty(\vb*{e}_z\times\vb*{E}\qty(t))\cdot \hat{\vb*{m}},
\end{align}
where $c$ is the speed of light and $\hat{\vb*{\mu}}$, $\hat{\vb*{m}}$ denote the electric and magnetic
dipole operators, respectively. 

In our study, we consider the electric field of a linearly chirped
circularly polarized pulse. 
The chirp is introduced by a quadratic spectral
phase function, which also leads to a quadratic term in the temporal phase
function and therefore to a linearly chirped pulse~\cite{Traeger_2012}. A
Taylor expansion around the center angular frequency $\omega_c$ results in
\begin{equation}
  \phi\qty(\omega)=\frac{1}{2}\left.\dv[2]{\phi\qty(\omega)}{\omega}\right|_{\omega=\omega_c}{\qty(\omega-\omega_c)}^2
  =\frac{1}{2}\mathrm{GDD}{\qty(\omega-\omega_c)}^2,
\end{equation}
where $\mathrm{GDD}$ is the group delay dispersion. For the case of left-circular polarization the pulse is then
given by~\cite{Traeger_2012}
\begin{align}
  \vb*{E}\qty(t)= \frac{E_0}{\gamma^{\frac{1}{4}}} e^{\frac{{\qty(t-t_c)}^2}{4\beta\gamma}} \mqty(\cos(\qty(\omega_c +a\qty(t-t_c))\qty(t-t_c)) \\ \sin(\qty(\omega_c +a\qty(t-t_c))\qty(t-t_c))),
  \label{eq:E-field}
\end{align}
with the field amplitude $E_0$, pulse center $t_c$ and
\begin{align}
  \beta=\frac{\Delta t^2}{8\ln(2)},\quad\gamma=1+\frac{\mathrm{GDD}^2}{4\beta^2},\quad a=\frac{\mathrm{GDD}}{8\beta^2\gamma},
\end{align}
where the pulse duration of the unchirped pulse $\Delta t$ is the full width half maximum of the corresponding intensity function.
The chirp increases the pulse length according to~\cite{Traeger_2012}
\begin{equation}
  \Delta t_{\mathrm{GDD}}=\sqrt{\Delta t^2+{\qty(4\ln(2)\frac{\mathrm{GDD}}{\Delta t})}^2},
\end{equation}
while keeping the pulse energy constant.

Throughout the range of employed $\mathrm{GDD}$ the resulting pulse lengths are of the
order of $100\,\mathrm{fs}$, while the time scale of rotations in 3MCP is in
the $\mathrm{ps}$ regime~\cite{li_microwave_1984}, cf.~\cref{sec:time_scale}.
We therefore neglect the rotational degree of freedom. Our
Hilbert space is then spanned by the relevant electronic-vibrational eigenstates of
$\hat{H}_0$.
While the experiment was performed at room temperature~\cite{das_control_2025},
we assume the initial state in its electronic and vibrational
ground state for simplicity.
This approximation is justified by the fact that most vibrational modes are nearly unpopulated at room temperature.
Specifically, by comparing with room-temperature calculations~\cite{kerber_anisotropy_2025} we have indeed observed that the main effect of considering the actual population of low-energy vibrational modes on the absorption spectra is an overall broadening.
In the following, we denote the electronic-vibrational ground
state by $\ket{0}$, the electronic state A with vibrational excitation $\vb*{v}$ by
$\ket{A\qty(\vb*{v})}$, and summarize the ten Rydberg states and their vibrational substructures as $\ket{Y}$.

The dynamics of the system depend on the specific molecular orientation and the
enantiomeric form. The orientation of the molecular frame with respect to the lab frame is parameterized by the Euler
angles $\qty(\alpha,\beta,\gamma)$ in the z-y-z convention.
We numerically confirmed that the system is almost perfectly rotationally symmetric around the $z$-axis, which is caused by the driving field containing sufficiently many cycles.
Specifically, a rotation about the pulse propagation axis results in a phase shift of the electric field in \cref{eq:E-field}.
For a sufficiently low chirp rate, such a phase shift corresponds approximately to a time translation of the optical carrier, of which the state populations are independent in the many-cycle case.
We can therefore omit the corresponding Euler angle in the following.

We obtain the lab-frame transition dipole moments by rotating the molecular-frame transition dipole moments via
\begin{align}
  \mel{\kappa\qty(\beta,\gamma)}{\hat{\vb*{\mu}}}{\eta\qty(\beta,\gamma)} &= R\qty(\beta,\gamma)\mel{\kappa}{\hat{\vb*{\mu}}}{\eta}, \notag\\
  \mel{\kappa\qty(\beta,\gamma)}{\hat{\vb*{m}}}{\eta\qty(\beta,\gamma)} &= R\qty(\beta,\gamma)\mel{\kappa}{\hat{\vb*{m}}}{\eta},
\end{align}
where $R\qty(\beta,\gamma)=R_z\qty(\alpha =0)R_y\qty(\beta)R_z\qty(\gamma)$ denotes the corresponding rotation matrix and $\ket{\kappa} ,\ket{\eta}$ denote the electronic-vibrational eigenstates of $\hat{H}_0$ in the molecular frame.
In our calculations, instead of changing the polarization of the light, we mirror the molecule. Both approaches are equivalent for an isotropic ensemble of molecular orientations.
We denote the respective enantiomer with subindices $R$ and $S$ in the following.

Starting from the electronic and vibrational ground state
$\ket{\psi\qty(\beta,\gamma,t=0)}_{R/S}=\ket{0(\beta,\gamma)}_{R/S}$
we calculate the final state
after excitation $\ket{\psi\qty(\beta,\gamma,t=T)}_{R/S}$ by solving the
time-dependent Schr\"odinger equation,
\begin{equation}
  i\hbar\frac{d\ket{\psi\qty(t)}}{dt}=\hat{H}{\qty(\beta,\gamma,t)}_{R/S}\ket{\psi\qty(t)}.
\end{equation}
From $\ket{\psi\qty(\beta,\gamma,t=T)}_{R/S}$, we find the populations in the Rydberg states $Y$ as
\begin{equation}
  n_{R/S}\qty(\beta,\gamma)=\sum_{Y} \abs{\braket{Y\qty(\beta,\gamma)}{\psi\qty(\beta,\gamma,t=T)}_{R/S}}^2.
\end{equation}
We calculate the anisotropy factor in terms of orientationally averaged
differential and mean enantiomeric populations $\Delta n$ and $\bar{n}$, that is,
we model the ion yield in our description by directly evaluating the total population in the Rydberg states at final time:
\begin{align}
  \Delta n =& \frac{1}{4\pi}\int_0^{2\pi}\int_0^{\pi}\sin(\beta)\mathrm{d}\beta\mathrm{d}\gamma \qty(n_{R}\qty(\beta ,\gamma)-n_{S}\qty(\beta,\gamma))\label{eq:delta_n} \\
  \bar{n} =& \frac{1}{4\pi}\int_0^{2\pi}\int_0^{\pi}\sin(\beta)\mathrm{d}\beta\mathrm{d}\gamma \frac{1}{2}\qty(n_{R}\qty(\beta ,\gamma)+n_{S}\qty(\beta,\gamma))\label{eq:n_bar} \\
  g&=\frac{\Delta n}{\bar{n}}.\label{eq:g}
\end{align}
At room temperature, 3MCP is present in the form of several conformers, with the equatorial and axial forms being predominant.
To account for this, we perform an additional averaging over the two conformers weighting their differential and mean absorptions by their respective proportions at room temperature $p$.
This yields the following expressions for the mean and differential populations as well as for the anisotropy factor,
\begin{align}
  \Delta n_{\mathrm{mix}}&=p_{\mathrm{eq}}\Delta n_{\mathrm{eq}}+p_{\mathrm{ax}}\Delta n_{\mathrm{ax}}\label{eq:delta_n_bar_c},\\
  \bar{n}_{\mathrm{mix}}&=p_{{\mathrm{eq}}}\bar{n}_{\mathrm{eq}}+p_{{\mathrm{ax}}}\bar{n}_{\mathrm{ax}} \label{eq:n_bar_bar_c},\\
  g_{\mathrm{mix}}&=\frac{\Delta n_{\mathrm{mix}}}{\bar{n}_{\mathrm{mix}}}.
  \label{eq:g_c}
\end{align}

\subsection{Models of the system}
The quantum-chemical data for the vibronic transitions between $\ket{0}$ and $\ket{A\qty(\vb*{v})}$, that is, the energetic positions
as well as electric and magnetic transition matrix elements are provided by the
computations within the time-independent framework of
Ref.~\cite{kerber_anisotropy_2025}.
We only include A-band transitions
within the range of approximately $301.7\,\mathrm{nm}$ to $318.6\,\mathrm{nm}$. We confirmed that including transitions outside of this range only has a negligible effect on the results.

Modeling of the second absorption step via quantum chemistry accounting for both the electronic and vibrational degrees of freedom is challenging.
Due to the numerous vibrational levels within the A band and the presence of several Rydberg states which can act as the final electronic state, it is difficult to unambiguously identify and characterize all relevant transitions.
Furthermore, several state-of-the-art approaches like time-dependent density functional theory or coupled-cluster methods are known to have significant difficulties in accurately describing Rydberg states~\cite{casida_molecular_1998,goetz_theoretical_2017,smydke_using_2023,creutzberg_computing_2023}.
To still be able to gain insight into the experimental observations, we employ two qualitative models designed to capture the essential properties of the system.
A ``minimal model'', which aims at understanding the basic mechanisms that
play a role in the process, and a ``randomized model'', which aims at reproducing
the qualitative features observed in the experiment.

For simplicity, the ionization step is not treated explicitly, as we assume the differential ion yield is primarily governed by the difference in total populations of the Rydberg states following the second excitation step according to \cref{eq:delta_n,eq:n_bar,eq:g}.
This is an approximation since the ion yield can, in principle, depend on which specific Rydberg state is occupied prior to ionization.

\subsubsection{Minimal model}\label{subsubsec:min_model}

In the minimal model, we restrict ourselves to the four most intense A band vibronic transitions of equatorial 3MCP.
The intensity of the transitions is quantified by their dipole strength,
\begin{equation}
  D_{A\qty(\vb*{v}) 0} = \abs{\mel{A\qty(\vb*{v})}{\hat{\vb*{\mu}}}{0}}^2
  + \frac{1}{c^2}\abs{\mel{A\qty(\vb*{v})}{\hat{\vb*{m}}}{0}}^2.\label{eq:dip_str}
\end{equation}
The Y band is represented by twice as many states as the A band and each A band state is independently coupled to two of the Y band states, which are energetically located
slightly above and below twice the energy of the respective A band state $E_{A\qty(\vb*{v})}$, that is,
\begin{equation}
  E^{A\qty(\vb*{v})\pm}_{Y}=2E_{A\qty(\vb*{v})} \pm \Delta E. 
  \label{eq:E_X}
\end{equation}
This construction reflects the fact that the energy required for the second excitation is unlikely to be identical to that for the first.
Each possible two-photon excitation path thereby reaches a distinct final Y band state, so by design, the minimal model excludes any coherent interference effects between different pathways.
This is motivated by the expectation that, in a more comprehensive model, the high density of states within both the A and Y bands would average out interference from individual contributions.
Finally, since the most significant transitions in the second excitation step are the most intense, we expect this excitation is predominantly electric in nature and therefore neglect magnetic transition moments.
Thus, the second excitation step does not introduce any additional anisotropy beyond that generated by the initial transition.
The corresponding electric transition matrix elements connecting each A band level to its two assigned Y band states are set uniformly to
\begin{equation}
  \mel{Y}{\hat{\mu}_i}{A\qty(\vb*{v})}=\mu_0 \qfor i\in{x,y,z}.
  \label{eq:mu_0_minmodel}
\end{equation}

\subsubsection{Randomized model}
In the randomized model, we include all A-band transitions, whose intensity according to their dipole strength, see \cref{eq:dip_str}, is at least $0.1\,\%$ compared to the strongest transition.
The Y band is represented by $N$ states, for which we randomly generate the energetic positions as well as the electric transition matrix elements from a uniform distribution according to
\begin{align}
  E_Y&\in\qty[E_{\min},E_{\max}]\qand \notag \\
  \mel{Y}{\hat{\mu}_i}{A\qty(\vb*{v})}&\in\qty[-\mu_{0},\mu_{0}].
  \label{eq:mu_0_rndmodel}
\end{align}
In contrast to the minimal model, interference between different excitation pathways is possible since each A band state is coupled to all Y band states.
This allows to confirm our assumptions from the minimal model that interference effects are negligible if our results prove to be mostly independent on the specific energetic positions and transition matrix elements, that is, the specific realization of our randomized model.

The randomized model is motivated by the assumption that only the specifics of the first transition are relevant, for which the associated anisotropy factor depends on the particular vibronic transition.
In contrast, the second absorption step is not expected to introduce additional anisotropy along any individual pathway.
Nevertheless, it is still anticipated to play a crucial role by enabling chirping to enhance or suppress contributions from pathways involving different vibronic A-band transitions, which would not be possible in a weak-field one-photon absorption scenario~\cite{shore_manipulating_2011}.
Accordingly, our random generation of Y band states serves this purpose without having to rely on detailed information about the second transition.

\section{Results}\label{sec:res}

All simulations were performed by numerically solving the time-dependent Schrödinger equation using the Chebyshev propagation scheme
implemented in the \emph{QDYN} library~\cite{qdyn}, with a time step of $\dd t
= 0.036\,\mathrm{fs}$. The orientational averaging in \Cref{eq:delta_n,eq:n_bar}
was carried out using the Lebedev--Laikov quadrature~\cite{lebedev1999quadrature}, 
which is particularly well suited for integrations over two Euler
angles~\cite{blech_numerical_2024}. A degree of $\mathcal{L}=5$
proved to yield sufficiently converged results.
For the unchirped pulse, a duration of $\Delta t = 50\,\mathrm{fs}$ was chosen.

The value (see \cref{eq:mu_0_minmodel}), respectively the upper limit (see \cref{eq:mu_0_rndmodel}) of the electric transition matrix elements between
A and Y band states is set to $\mu_0=0.08ea_0$.
This choice is motivated by the assumption, that the most intense transitions of the second exitation step are dipole-allowed together with
experimental findings indicating that the maximum
intensity within the weak A band of 3MCP and related chiral ketones is roughly
two orders of magnitude smaller than that of the higher-lying dipole-allowed
electronic transitions~\cite{feinleib_vapour-phase_1968,pulm_theoretical_1997}.

In selecting the peak intensity, we are guided by the experiment of Ref.~\cite{das_control_2025}, where the anisotropy is found to be largely insensitive to the applied intensities.
Since we neglect ionization and the associated depopulation of Rydberg states into the continuum, we set $I_0 = 10^{10}\,\mathrm{W/cm^2}$, two orders of magnitude below the experimental value.
Nevertheless, our model reproduces the robustness with respect to the laser intensity observed in the experiment as increasing its value by an order of magnitude has only minor effects on our results. 

\subsection{Minimal Model}\label{subsec:min_model}
We begin by developing a basic understanding of the system with our minimal model.
The four most important transitions of the A band of equatorial 3MCP according to their dipole strengths are summarized in \cref{tab:min_model}.
\begin{table}[tb]
    \centering
    \caption{Most relevant transitions of the A band of equatorial 3MCP according to their dipole strengths.}\label{tab:min_model}
    \begin{tabular}{l c c c c}
        \toprule
                               & $\lambda$ $\qty(\mathrm{nm})$ & $D$ $\qty(10^{-4} e^2 a_0^2)$ & $g$   \\
        \midrule
        $A\qty(\vb*{v}_{1})\leftarrow 0$ & 309.69 & 1.57 & 0.347 \\
        $A\qty(\vb*{v}_{2})\leftarrow 0$ & 309.42 & 2.15 & 0.367 \\
        $A\qty(\vb*{v}_{3})\leftarrow 0$ & 307.47 & 0.63 & 0.003 \\
        $A\qty(\vb*{v}_{4})\leftarrow 0$ & 307.21 & 0.98 & 0.003 \\
        \bottomrule
    \end{tabular}
\end{table}
Accordingly, the A band in the minimal model comprises two intense transitions near
$309\,\mathrm{nm}$ with anisotropy factors of approximately $g \approx 0.35$,
and two weaker transitions around $307\,\mathrm{nm}$ characterized by much
smaller anisotropy factors of about $g \approx 0.003$.
These anisotropy factors associated with individual A-band transitions were calculated from the corresponding rotatory strengths $R_{A\qty(\vb*{v}) 0}$ and dipole strengths (\cref{eq:dip_str})~\cite{CDnakanishi,kerber_anisotropy_2025},
\begin{align}
  R_{A\qty(\vb*{v}) 0} &= \Im{\mel{0}{\hat{\vb*{\mu}}}{A\qty(\vb*{v})}\cdot\mel{A\qty(\vb*{v})}{\hat{\vb*{m}}}{0}} \notag\\
  g\qty(A\qty(\vb*{v})) &= \frac{\flatfrac{4R_{A\qty(\vb*{v}) 0}}{c}}{D_{A\qty(\vb*{v}) 0}}.
\end{align}
The energy shift of \cref{eq:E_X} is chosen as $\Delta E = 0.0005 E_\mathrm{h}\approx 0.0136\,\mathrm{eV}$, which corresponds approximately to the lower end of vibrational excitation energies in the A band of 3MCP~\cite{kerber_anisotropy_2025}, such that we achieve a plausible level spacing.

\begin{figure}[bt]
  \centerline{\includegraphics[width=8.6cm]{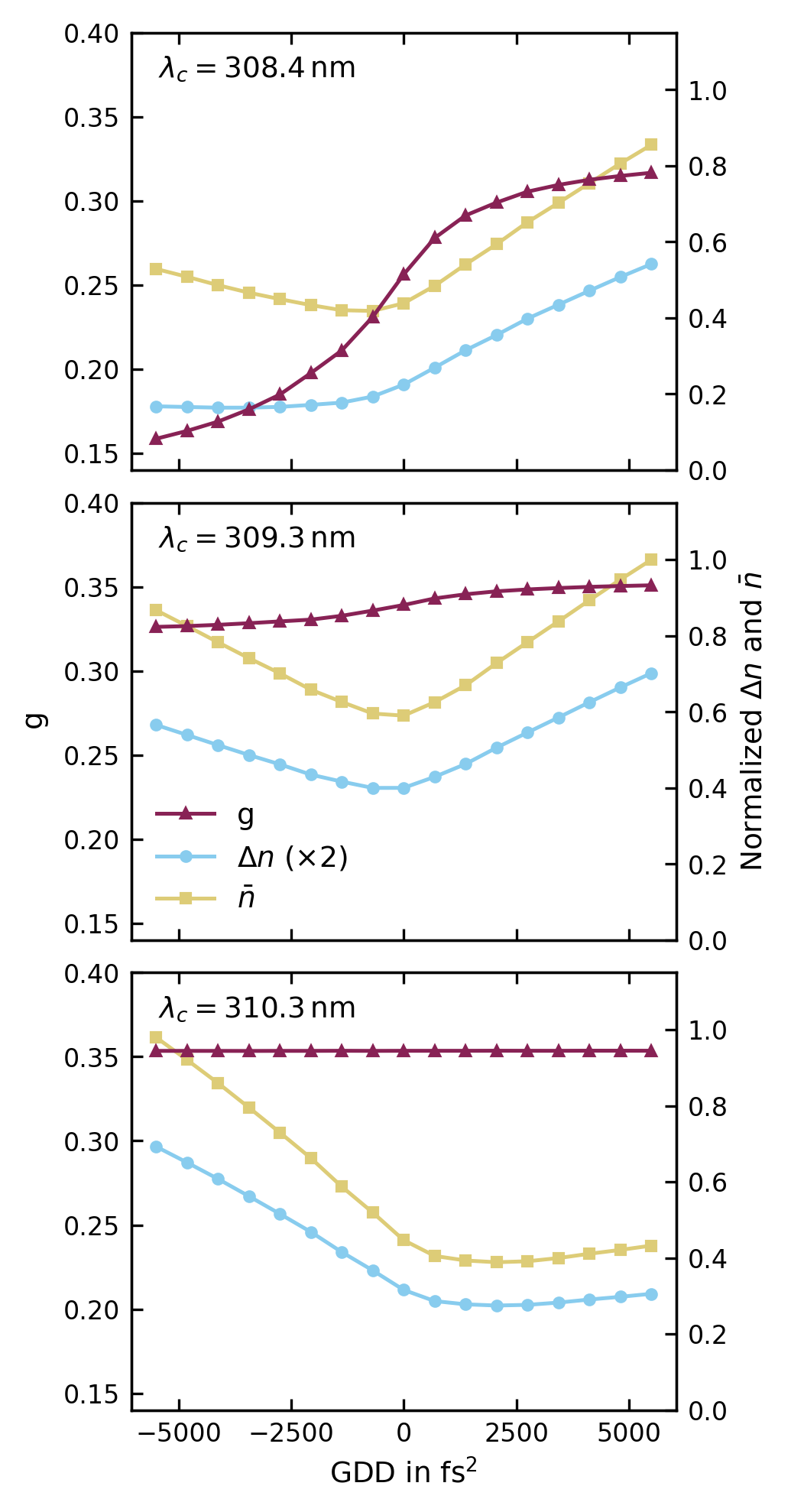}}
  \caption{Results for the ``minimal model''.
    Dependence of the anisotropy factor ($g$, left axis), CD and ABS ($\Delta n$ and $\bar{n}$, right axis) on the applied GDD for three different central wavelengths of the linearly chirped laser pulse.
  }\label{fig:model4n8}
\end{figure}

\Cref{fig:model4n8} displays the anisotropy factor $g$ (red curve) according to \cref{eq:g} as a
function of the group delay dispersion for three different central
wavelengths $\lambda_c$ of the driving laser pulse.
For $\lambda_c = 310.3\,\mathrm{nm}$, the anisotropy factor remains close to
$0.35$, independent of the applied GDD.\@ This value coincides with the
anisotropy of the two intense transitions. The reason is that the central
wavelength lies energetically below these strong transitions, while the weaker
transitions with small anisotropy factors are too far detuned to contribute
significantly. Consequently, all relevant excitation pathways share the same
anisotropy, resulting in a constant overall value.

In contrast, the mean and differential absorption (yellow and blue curves) increase for negative GDD.\@
A negative GDD implies that lower frequencies are delayed relative to higher frequencies within the pulse.
Since the central frequency is below the energies of the intense transitions, the higher-frequency components are responsible for driving the first excitation step.
As a result, these components efficiently populate the A band states before the subsequent arrival of lower-frequency components, which then drive the second excitation step.

\begin{figure}[bt]
  \centerline{\includegraphics[width=8.6cm]{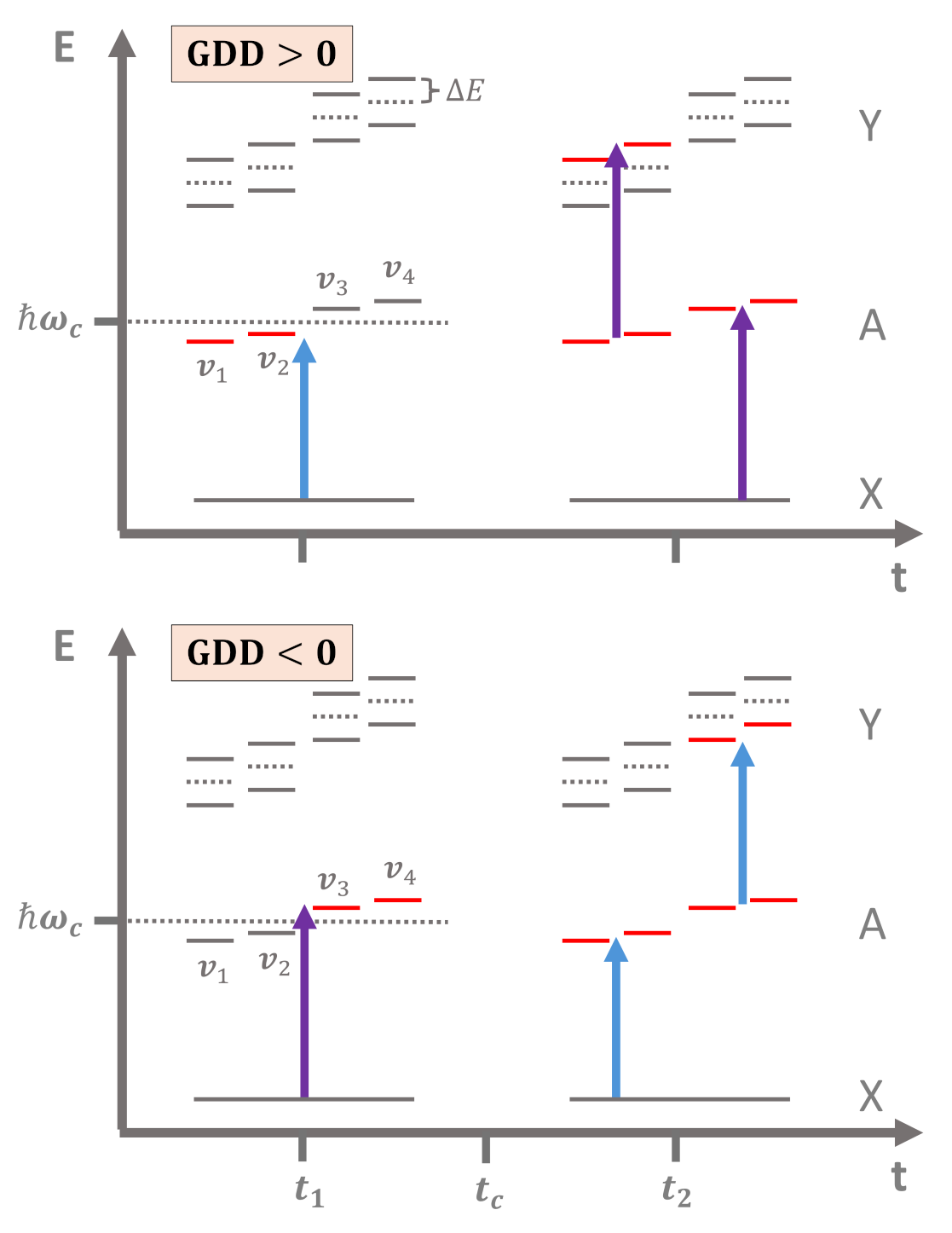}}
  \caption{
    Schematic illustration of the preferential enhancement of pathways for the case of a central pulse frequency ($\omega_c$) which energetically lies in between the two intense ($A\qty(\vb*{v}_1)$, $A\qty(\vb*{v}_2)$) and the two weak transitions ($A\qty(\vb*{v}_3)$, $A\qty(\vb*{v}_4)$), as in the upper two panels of \cref{fig:model4n8}.
    States in the same column are coupled with each other: each A band state (at energy $E_{A\qty(\vb*{v})}$) is coupled to two Y band states at energies $2E_{A\qty(\vb*{v})}\pm \Delta E$.
    The excitation process is shown at two points in time $t_1$ and $t_2$, before and after the pulse center $t_c$.
    For positive chirp ($\mathrm{GDD}>0$), low frequencies (blue), resonant with the intense transitions, arrive first, enhancing subsequent population (red bars) of the corresponding upper Y band states via photon energies $\hbar\omega_{t_2}>E_{A\qty(\vb*{v})}$.
    For negative chirp ($\mathrm{GDD}<0$), high frequencies (purple), resonant with the weak transitions, arrive first, enhancing subsequent population of the corresponding lower Y band states via photon energies $\hbar\omega_{t_2}<E_{A\qty(\vb*{v})}$.
  }\label{fig:mechanism}
\end{figure}

When the central wavelength is shifted to higher energies (upper two panels in \cref{fig:model4n8}), that is, closer to resonance with the
weak transitions, the overall anisotropy factor decreases. This trend reflects
the growing contribution of the weak transitions, which are characterized by
small anisotropy values.
In addition, the applied central wavelengths ($309.3\,\mathrm{nm}$ and $308.4\,\mathrm{nm}$) lie
energetically above the intense transitions but below the weak transitions. 
Under
these conditions, as illustrated in \cref{fig:mechanism}, negative GDD enhances the contribution of the weak
transitions via efficient population by the mechanism discussed above, whereas positive GDD
preferentially strengthens the intense transitions.
As a result, the anisotropy
factor increases from small values at negative GDD to larger values at positive
GDD.\@
This is particularly prominent for the $\lambda_c=308.4\,\mathrm{nm}$ case.
For negative GDD, CD remains nearly constant while ABS increases, mirroring the small anisotropy factors of the weak transitions.
In contrast, for positive GDD both CD and ABS increase, resembling the large anisotropy factors of the intense transitions.

The minimal model demonstrates that the second absorption step is essential to explain the asymmetry of the anisotropy with respect to the sign of the GDD.\@
Pathways through A-band transitions that are resonant with the pulse’s early frequencies are preferentially enhanced, which is in line with the sequential resonance mechanism~\cite{cao_intrapulse_1998,yakovlev_chirped_1998,balling_interference_1994}.

\subsection{Randomized model}

In the randomized model, we randomly generate $N$ states in the Rydberg band.
We tested various $N \leq 400$ and found the results are robust in terms of the number of generated states, provided $N\geq 150$, see \cref{sec:rndX}.
In the following, we employ $N=270$.
The energetic positions of the Y band states are drawn from a uniform distribution among the domain $E_Y\in\qty[7.782\,\mathrm{eV},8.218\,\mathrm{eV}]$, which is centered on twice the energy of a photon of $310\,\mathrm{nm}$ wavelength.
Extending the domain to include energies outside of this interval only negligibly alters the results.
We repeat our simulations for ten sets of randomized energies and transition moments in order to investigate the sensitivity of our results to these parameters.

\begin{figure}[bt]
  \centerline{\includegraphics[width=8.6cm]{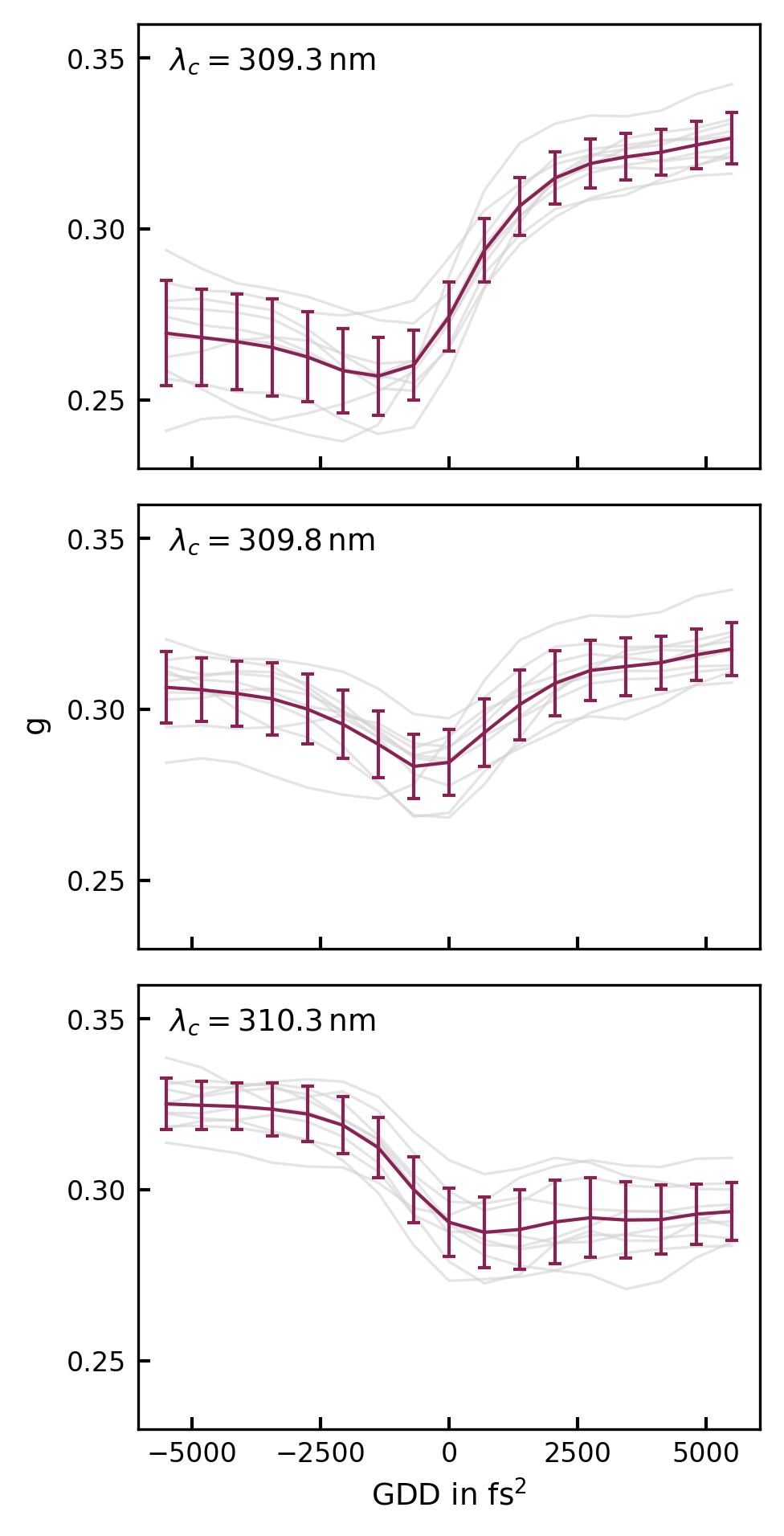}}
  \caption{Results for the ``randomized model''.
    Dependence of the anisotropy factor on the applied GDD for three different central wavelengths of the linearly chirped laser pulse for the equatorial conformer of 3MCP.
  }\label{fig:full_eq3mcp}
\end{figure}

\Cref{fig:full_eq3mcp} shows the results for the equatorial conformer of 3MCP.\@
The graphs represent the mean anisotropy factor of the calculations employing different randomly generated data sets.
The standard deviation for each GDD value is included as error bars.
The outcomes of the individual runs are included as pale lines.
As before, the results are shown for three different central wavelengths of the laser.

Before discussing our results, we list the main features of Ref.~\cite{das_control_2025}:
(1) Relative to the bandwidth-limited case at $\mathrm{GDD}=0$, the anisotropy factor increases for both positive and negative chirping of the pulse.
(2) The magnitude of the anisotropy factor is roughly between $2.5\,\%$ and $10\,\%$ among all measurements.
(3) This enhancement of anisotropy is asymmetric with respect to the chirp sign.
(4) The asymmetry depends on the central wavelength: at lower central wavelengths, positive chirp leads to a stronger enhancement than negative chirp, while at higher wavelengths this relationship is inverted.

We now analyse the results shown in \cref{fig:full_eq3mcp} in light of these features. 
The mean value of the anisotropy factor shows the same qualitative dependence on the GDD and the central wavelength as found in Ref.~\cite{das_control_2025}.
With our particular choice of central wavelengths the reversal of the asymmetry with respect to the chirp sign is captured.
For $\lambda_c=309.3\,\mathrm{nm}$, the anisotropy factor is maximal for the largest positive value of the GDD at $5500\,\mathrm{fs}^2$.
From there on it decreases to a minimum at around zero chirp and then increases again for negative GDD to a local maximum at $\mathrm{GDD}=-5500\,\mathrm{fs}^2$.
The behaviour is exactly opposite, when increasing the central wavelength to $\lambda_c=310.3\,\mathrm{nm}$.
While the dip around zero chirp is retained, now the anisotropy factor is largest for negative GDD.
Remarkably, the qualitative agreement with the experimental findings~\cite{das_control_2025} is independent of the specific random realization of the model, which is highlighted by the error bars.
The minimum anisotropy factor found in our theoretical simulations is, however, not at exactly zero GDD, but in its vicinity which we rationalize with the qualitative nature of our model.

\begin{figure}[bt]
  \centerline{\includegraphics[width=8.6cm]{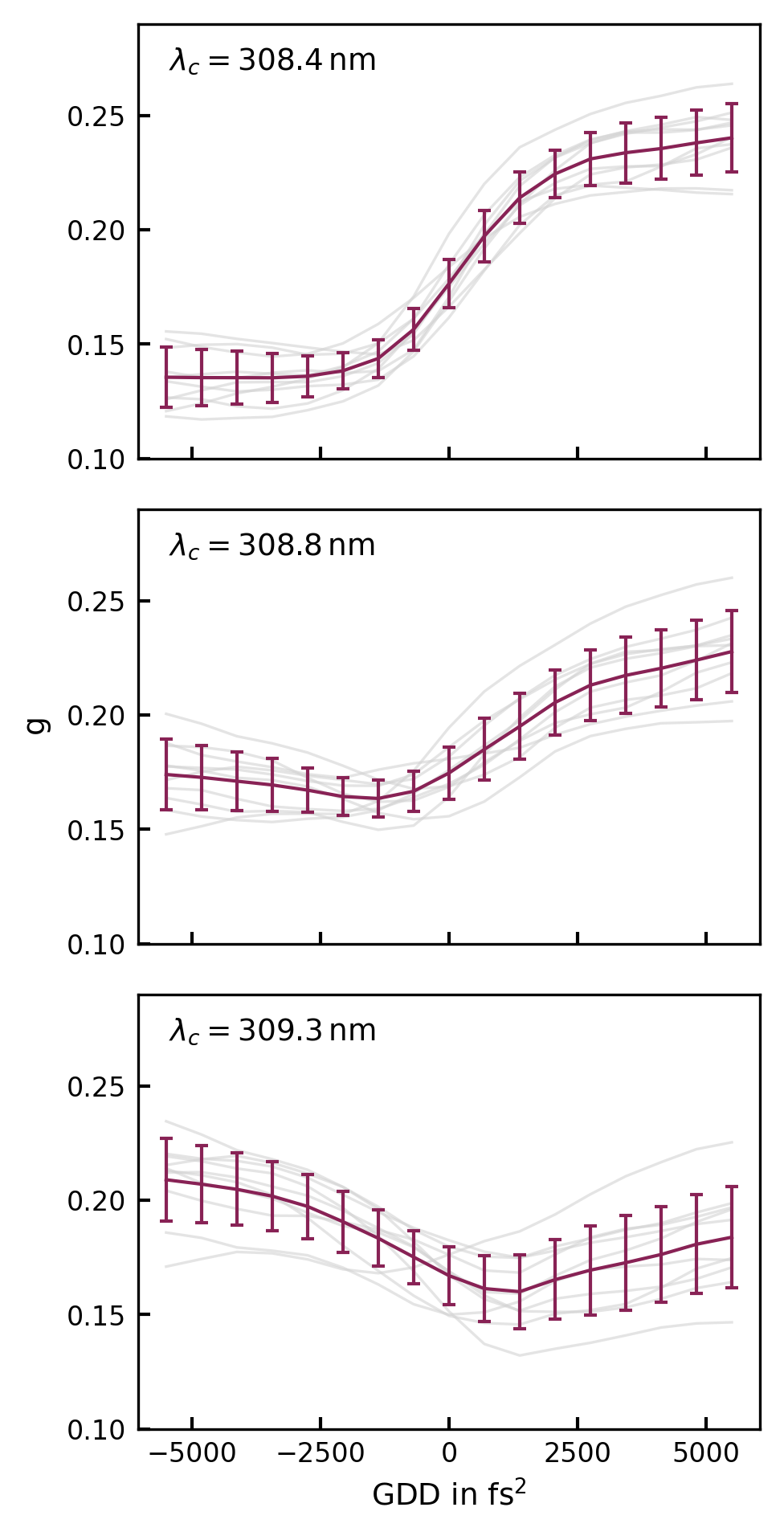}}
  \caption{Results for the ``randomized model''.
    Dependence of the anisotropy factor on the applied GDD for three different central wavelengths of the linearly chirped laser pulse for 3MCP.
    The results correspond to a mixture of 78:22 equatorial and axial conformers as in Ref.~\cite{kerber_anisotropy_2025}.
  }\label{fig:full_mix}
\end{figure}

The anisotropy factors in \cref{fig:full_eq3mcp} are much larger than the experimental values. This is caused by the corresponding simulations only involving the equatorial conformer and thus neglecting the presence of the axial conformer of 3MCP at room temperature.
In a next step we therefore add the axial conformer and average according to \cref{eq:delta_n_bar_c,eq:n_bar_bar_c,eq:g_c}.
The ratio of conformers is chosen identically to Ref.~\cite{kerber_anisotropy_2025}, i.e.~$78\,\%$ equatorial to $22\,\%$ axial 3MCP~\cite{he_determining_2004,al-basheer_spectroscopic_2007}.
\Cref{fig:full_mix} shows the results for the conformational mixture.
We once again show the calculations associated with three different central wavelengths of the laser field, such that the reversal of the asymmetry is captured.
The dip at around zero chirp for the mean anisotropy factor is much less pronounced than before.
Nevertheless, the reversal of the asymmetry of the anisotropy with respect to positive and negative values of the GDD at around $309\,\mathrm{nm}$ persists.
Furthermore the overall magnitude of the anisotropy factor varies from about $0.15$ to $0.25$ which is much closer to the experimental values.

We conclude by emphasizing once again that the results are largely independent of the specific random realization.
This supports our assumption that coherence effects are indeed negligible and that control is primarily achieved through preferential pathway enhancement via the sequential resonance mechanism, as already elucidated by the minimal model.

\section{Conclusions}\label{sec:conclusions}

We have conducted a theoretical investigation of circular dichroism following $1+1+1$ resonance-enhanced multiphoton ionization of 3-methylcyclopentanone, focusing in particular on understanding the mechanism behind the control of the anisotropy factor using linearly chirped pulses.

In contrast to previous theoretical models~\cite{horsch_circular_2011,kroner_chiral_2011,mondelo-martell_increasing_2022,mondelo-martell_correction_2023} we have established the first theoretical framework that incorporates both the non-Condon vibrational structure of the A band and the second absorption step.
Both elements have proven to be crucial for understanding how the anisotropy of the ion yield can be modulated via linearly chirped pulses.

Our investigation unraveled two key factors in this context.
First, the pathway dependence of the anisotropy factor, originating from the weak nature of the first electronic transition.
Second, a preferential enhancement of certain pathways depending on the applied group delay dispersion, resulting from the inclusion of the second absorption step, which enables a sequential resonance mechanism.

While state-of-the-art quantum-chemical results~\cite{kerber_anisotropy_2025} were employed for the first absorption step, the parameters for the second transition were generated randomly.
This effective modeling approach was chosen because a full quantum-chemical treatment of the second absorption step is challenging: it would require describing transitions from numerous vibrational A‑band levels to the vibrational substructure of ten different Rydberg states.
As a result of our modeling, although we have not yet achieved accurate quantitative agreement, we were able to provide convincing qualitative explanations for the experimental observations~\cite{das_control_2025}.

A natural refinement of the current model would be an improved description of the second absorption step, in particular through accurate quantum-chemical calculations.
In addition, the model could be extended by explicitly incorporating the ionization step. 
In this context, multiphoton ionization proceeding directly via the Rydberg states of ketones may offer additional opportunities for control.
On the one hand, these transitions are associated with small anisotropy factors due to their electric-dipole character~\cite{pulm_theoretical_1997}.
On the other hand, the Rydberg states and the corresponding ionic states are known to share a similar vibrational structure~\cite{kastner_high-resolution_2020,singh_experimental_2020}.
Ionization pathways involving different intermediate Rydberg states therefore lead to the same final-state vibrational excitations, thereby allowing for coherent interference.
Such interferences could, in principle, be controlled in a pump--probe scheme by preparing a superposition of intermediate Rydberg states and ionizing it after a controllable delay.

Finally, low-temperature measurements could allow to gain further insight into conformational dynamics.
Moreover, photoelectron--photoion coincidence measurements are currently being developed that will allow for resolving the CDIY for indivdual Rydberg states~\cite{personal_comm}.
By investigating if and in how far the results of such measurements are reproduced by theoretical models, we expect to further improve our understanding of the role of the second absorption and ionization steps.

\section{Acknowledgements}
We are grateful to Oliver Kreuz for his support with the quantum chemistry and to Hendrike Braun, Sagnik Das and Monika Leibscher for fruitful discussions and feedback.
Furthermore, we would like to thank the HPC Service of the physics department and FUB-IT, Freie Universit\"at Berlin, for computing time.
This work is funded by the Deutsche Forschungsgemeinschaft (DFG, German Research Foundation) -- Projektnummer 328961117 -- SFB ELCH 1319.

\bibliography{./bib/chirpCD.bib}

\appendix

\section{Time scale of the rotational degree of freedom}\label{sec:time_scale}

The characteristic time scale for 3MCP can be estimated using both classical and quantum mechanical methods. 
While the classical method provides a clear illustration, the quantum approach is more rigorous. Still, both methods yield consistent results for the rotational time scale.
We first derive expressions for the two time scales $T_L^{\mathrm{classical}}$ and $T_L^{\mathrm{QM}}$, then calculate both for a representative choice of $L$.

In the rigid rotor model, 3MCP is treated as an asymmetric top, which is characterized by three distinct rotational constants.
For the purposes of estimation, however, we simplify 3MCP to a linear rotor described by a single rotational constant.
Specifically, we select the largest rotational constant, $A \approx 5423\,\mathrm{MHz}\times h$~\cite{li_microwave_1984}, as this corresponds to the smallest, and thus most conservative, time scale.
The associated moment of inertia is given by
\begin{equation}
  I_A = \frac{\hbar^2}{2A}.
\end{equation}

The classical time scale corresponds to the period of a full classical rotation.
To determine this, we first calculate the rotational energy for a linear rotor in the quantum state characterized by rotational quantum number $L$,
\begin{equation}
  E_L = A L\qty(L + 1).
\end{equation}
The relationship between the rotational energy and the classical period of rotation, $T^{\mathrm{classical}}_L$, is then given by
\begin{align}
  E_L &= \frac{1}{2}I_A {\qty(\frac{2\pi}{T^{\mathrm{classical}}_L})}^2, \notag \\
  T_L^{\mathrm{classical}} &= 2\pi \sqrt{\frac{I_A}{2 E_L}} = \frac{h}{2A} \frac{1}{\sqrt{L\qty(L+1)}}.
\end{align}

In contrast, the quantum time scale is determined by the characteristic energy spacing between rotational states.
For a given electronic-vibrational transition, the total transition energy includes the contribution from the rotational part.
The rotational transition can occur from a state with quantum number $L$ to either $L+1$ or $L-1$ with an energy difference of $E_{L+1} - E_{L-1}$ between the two final states.
If this difference is small compared to the laser bandwidth, the laser cannot distinguish between the final states, allowing the rotational degree of freedom to be neglected.
Equivalently, the quantum mechanical time scale associated with this energy splitting can be defined by the temporal evolution of the relative phase, $e^{-\iu \frac{\qty(E_{L+1} - E_{L-1})}{\hbar}t}$.
Thus, the rotational degree of freedom is negligible as long as the pulse duration is much shorter than
\begin{equation}
  T^{\mathrm{qm}}_L = \frac{h}{E_{L+1} - E_{L-1}} = \frac{h}{2A}\frac{1}{\qty(2L + 1)},
\end{equation}
where any phase difference becomes insignificant.

For large values of $L$, the quantum and classical estimates for the rotational time scale become comparable, with $T_L^{\mathrm{qm}}\approx \frac{1}{2}T_L^{\mathrm{classical}}$.
As $L$ increases, the time scale decreases, making the rotational degree of freedom more relevant.
We estimate a plausible $L$ using a thermal distribution over rotational states.
The Boltzmann population of the rotational state with quantum number $L$ relative to the ground state is
\begin{equation}
  p_L = \qty(2L + 1) e^{-\frac{E_L}{k_B T}},
\end{equation}
where $k_B$ is the Boltzmann constant and $\qty(2L + 1)$ accounts for the degeneracy of the $L$ state.
At room temperature ($T = 300K$) the maximal population in our simplified linear rigid rotor model of 3MCP occurs at $L_{0} = 24$.
The corresponding time scales are
\begin{align}
  T_{L_{0}}^{\mathrm{classical}} &= 3.76\,\mathrm{ps}, \notag \\
  T^{\mathrm{qm}}_{L_{0}} &= 1.88\,\mathrm{ps}.
\end{align}

\section{Robustness of the randomized model with respect to the number of states in the Rydberg band}\label{sec:rndX}

\begin{figure}[htbp]
  \centerline{\includegraphics[width=8.6cm]{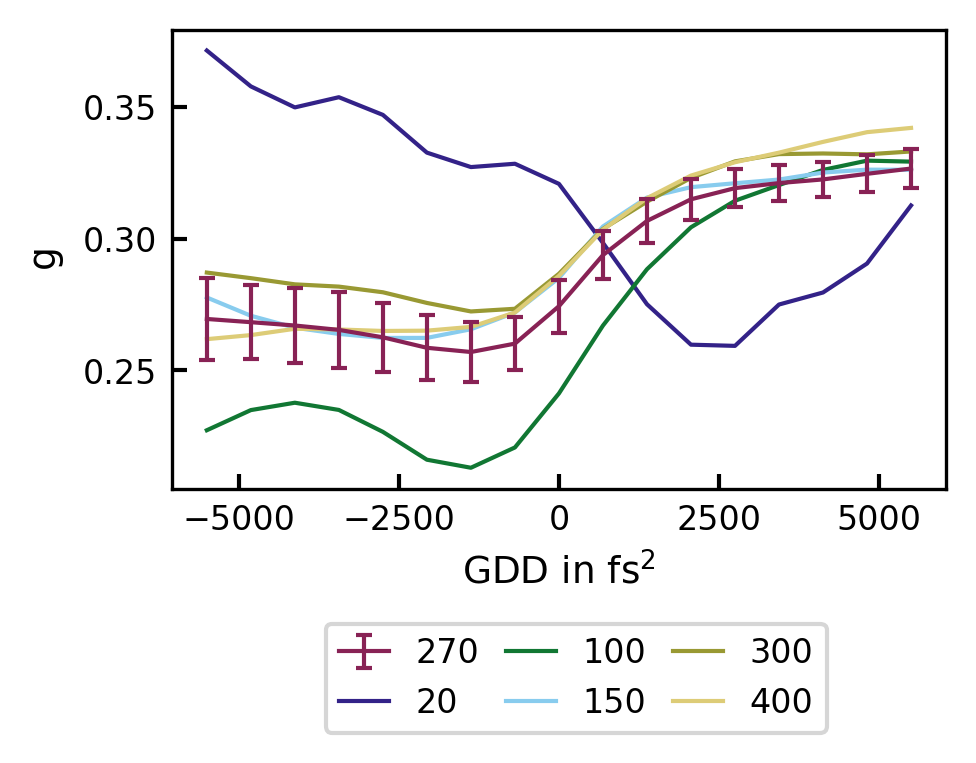}}
  \caption{Comparison of anisotropy factors between the averaged result for $N=270$ from \cref{fig:full_eq3mcp} and individual calculations performed with $N\in\qty{20,100,150,200,300,400}$. All results are shown for a central wavelength of $\lambda_c = 309.3\,\mathrm{nm}$.
  }\label{fig:rndX}
\end{figure}

\Cref{fig:rndX} compares the averaged result of \cref{fig:full_eq3mcp} with individual simulations using different numbers of random states in the Rydberg band at $\lambda_c = 309.3\,\mathrm{nm}$, specifically $N\in\qty{20,100,150,300,400}$.
The results are robust for $N\geq 150$, whereas significant qualitative differences appear for $N\leq 100$.
Nevertheless, we could observe that the reversal of the asymmetry persists down to $N = 50$.

\end{document}